\documentclass[aps,prb,twocolumn,showpacs]{revtex4}

\usepackage{graphicx}

\begin{document}

\title{Performance of a bipolar single electron device}

\author{Heinz--Olaf M{\"u}ller}
\affiliation{Integrated Systems Engineering AG, Balgriststrasse 102, 
  CH--8008 Z{\"u}rich, Switzerland}
\email{hom@ise.ch}

\date{\today}

\begin{abstract}
  A small scale bipolar transistor with polysilicon emitter will, depending on
  the emitter window size, display suppression of the hole transport due to
  single electron effects. In this paper the resulting base current
  suppression is computed in terms of the orthodox theory of single electron
  tunneling and a recombination time approximation. The possible application
  of the transistor as readout system for Coulomb blockade device circuits is
  discussed.
\end{abstract}


\maketitle

{\em Introduction --}\/ The appeal of bipolar transistors is their appreciable
current drive $\beta = I_c/I_b > 100$. The use of polysilicon (poly--Si)
emitters has helped to maintain this figure of merit despite increasing
challenges from continuous down--scaling~\cite{kap1}. The effect of the new
material is attributed to the poly--Si grain boundaries obstructing the hole
transport thus reducing $I_b$ and increasing $\beta$~\cite{yur1,rin1}. Point
contact measurements of a few poly--Si grains resulted in effective grain
boundary energy barriers of up to 80\,meV which translates into three times
room temperature. These findings correspond roughly to the values found in
molecular dynamics simulations~\cite{ber3}.

This work studies the performance of radically scaled bipolar transistors
with small emitter windows, especially the hole transport in the poly--Si
emitter. For these devices single electron effects can be expected, {\em i.e.}
charged grains prevent subsequent holes from entering. Thus the hole current
is further reduced and the current drive $\beta$ enhanced.


The studied device is similar to the tunnel emitter
transistor~\cite{chu1,gre1} insofar as it uses tunnel barriers in the emitter
to reduce the base current. Otherwise the two devices are very dissimilar: the
tunnel emitter transistor employs a metal emitter that can electrically induce
the transistor's base region.

The studied bipolar transistor differs fundamentally from the single hole
transistor~\cite{doe1}. The former is a rather classical device involving a
mesoscopic effect for the enhancement of its operation while the latter is a
true mesoscopic device similar to the single electron transistor~\cite{ful2}.

Single electron effects in poly--Si have been reported since
1994~\cite{yan2,yan8}. Single grains of a very thin (less than 10\,nm) undoped
poly--Si films could be charged by applying a large gate bias (44\ldots60\,V).
This charging resulted in a noticeable hysteresis of the $I_d$--$V_{gs}$
characteristics. A grain capacitance of $\sim2$\,aF was deduced from the
hysteresis.

Heavily doped poly--Si nanowires ($5\times10^{19}\,{\rm cm}^{-3}$) of larger
cross section ($20\times30\,{\rm nm}^2$) showed single electron effects at low
temperature (4.2\,K)~\cite{irv1,tan1}. Again, the grain capacitance as deduced
from electrical measurements was $\sim2$\,aF while the average grain size is
estimated to $\sim20$\,nm.


\begin{figure}
\ifx\pdfoutput\undefined
  \includegraphics[width=3.4in]{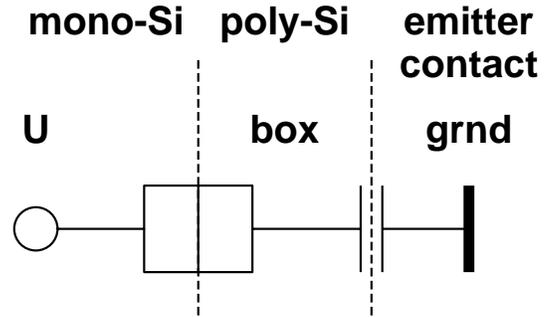}
\else
  \includegraphics[width=3.4in]{figure1.pdf}
\fi
\caption{Schematic illustration of the single electron box model used in this
  paper. The transistor base is to the left, the emitter contact to the right.
  An injection voltage $U$ forces holes into the poly--Si emitter, {\em i.e.}
  the single electron box, where the holes fall victim to recombination.}
\label{fig0}
\end{figure}

{\em Method --}\/ The simulation of the poly--Si emitter must feature charging
effects for the device under consideration. To this end the model of a single
electron box~\cite{laf4,est1}, see Fig.~\ref{fig0}, is used. The box
represents a single grain of the emitter. It lives between a tunnel junction
(base side) and a capacitance (emitter contact side). This model is a stark
simplification of the real poly--Si emitter insofar as it excludes hole
transport into the emitter contact and assumes only one grain wide emitters.
The first assumption is often close to the truth since the emitter width is
designed to exceed the hole diffusion length. As for the second
simplification, a single electron trap model~\cite{mat4} might be more
appropriate, but its simulation is numerically more involved and its behavior
does not qualitatively deviate from that of a single electron box in the
current context.

The state of a single electron system in an orthodox situation is given in
terms of the charge number states of each of its island electrodes, which form
good quantum numbers~\cite{lik5}. The single electron box consists of one
island and the number of excess charges $n$ on this island suffices for the
description.

In the original model of the single electron box~\cite{laf4,est1} there is no
average net current in or out. In the case of the poly--Si emitter the hole
current arises due to recombination. Therefore a recombination model is added
to the box description. A constant recombination time $\tau_{\rm rec}$ is
assumed leading to a recombination rate of a state $n$
\begin{equation}
\label{grec}
\Gamma_{\rm rec}[n\to n-{\rm sgn}(n)] = \frac{|n|}{\tau_{\rm rec}},
\end{equation}
$\Gamma[n\to m]=0$ for $m\neq n-{\rm sgn}(n)$. 

\begin{figure}
\ifx\pdfoutput\undefined
  \includegraphics[width=3.4in]{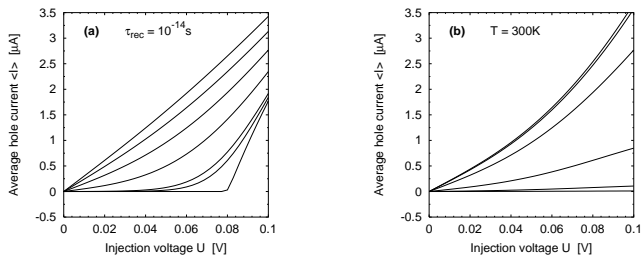}
\else
  \includegraphics[width=3.4in]{figure2.pdf}
\fi
\caption{Hole current into the poly--Si emitter as function of the injection
  voltage $U$. The box parameters are $C_g = C = 1\,$aF, $R = 10\,{\rm
    k}\Omega$.  (a) Temperature dependence, the data correspond to 4.2\,K,
  77\,K, 100\,K, 200\,K, 300\,K, 400\,K, and 500\,K (from bottom to top),
  $\tau_{\rm rec} = 10^{-14}\,$s. (b) $\tau_{\rm rec}$ dependence, the data
  correspond to $10^{-11}\,$s, $10^{-12}\,$s ,$10^{-13}\,$s, $10^{-14}\,$s,
  $10^{-15}\,$s, $10^{-16}\,$s (from bottom to top), $T = 300\,$K.}
\label{fig1}
\end{figure}

The recombination rate $\Gamma_{\rm rec}[n\to n\pm1]$ is added to the
corresponding single electron tunneling rate $\Gamma_{\rm set}[n\to n\pm1]$
which is extensively discussed in the literature~\cite{ing2}
\begin{displaymath}
  \Gamma_{\rm tot}[n\to n\pm1] = \Gamma_{\rm rec}[n\to n\pm1]+
  \Gamma_{\rm set}[n\to n\pm1].
\end{displaymath}
The total rates $\Gamma_{\rm tot}[n\to n\pm1]$ are used to set up an orthodox
master equation for the evolution of probability of the state $n$, ${\rm
  d}p(n,t)/{\rm d}t$~\cite{ave3}. This equation possesses a well--known
stationary solution~\cite{amm2,seu1}, $p(n)$, which holds for processes slow
compared to $1/\Gamma_{\rm tot}$. It is used in the current context to express
the stationary average injection current $\langle I\rangle$, which is plotted
in Fig.~\ref{fig1}.

{\em Discussion --}\/ Which are the geometric requirements for the observation
of the discussed effect? The emitter structure corresponds to a vertical setup
of the poly--Si films discussed before. In the foreseeable future it will be
impossible to use films of the dimension of Refs.~\onlinecite{yan2,yan8}.
However, condition are more relaxed for the application under discussion
because retention time is no issue. Structures of the type of
Refs.~\onlinecite{irv1,tan1} suffice in the current context and an emitter
window of $50\times50\,{\rm nm}^2$ appears likely to be required.

The apparatus of the orthodox theory applies only to systems satisfying
$R>\hbar/e^2\approx6.45\,{\rm k}\Omega$. Otherwise, $n$ is no good quantum
number anymore and quantum fluctuations lead to a larger hole
current~\cite{wan2}. The grain boundary resistance of a poly--Si film, like
other material parameters, depends strongly on the process conditions. However
following Ref.~\onlinecite{fur3}, it can be purported that the orthodox theory
is applicable in certain instances.

The injection voltage $U$ can be expressed by means of Boltzmann statistics in
the low injection approximation, $U = k_B T/e\approx26\,$mV for room
temperature. In contrast, the number of injected charges is given by the base
voltage $V_{be}$.

The hole currents of Fig.~\ref{fig1} can be set in relation to the
recombination current without Coulomb blockade, $e/(R\,C+\tau_{\rm
  rec})\approx10^{-5}\,$A. Therefore an injection reduction to $0.1$ can be
expected for the given parameters. Conventionally $\tau_{\rm rec}$ is assumed
to exceed the assumed value $10^{-14}\,$s considerably thus furthering the
performance enhancement of the Coulomb blockade emitter.

Besides injection into the emitter (and subsequent recombination), holes
recombine in the quasi-neutral base and the emitter--base space charge
region. The influence of the different recombination processes on the current
drive $\beta$ is conveniently by the partial values $\beta_1$ (quasi-neutral
base), $\beta_2$ (space charge region), and $\beta_3$ (emitter injection),
\begin{displaymath}
\beta = \frac{1}{1/\beta_1+1/\beta_2+1/\beta_3}.
\end{displaymath}
Typical values are $\beta_1\approx10^6$ and $\beta_2\approx10^3$ for modern
devices so that $\beta_{1,2}\gg\beta_3$ and $\beta\approx\beta_3$. The
performance gain of Coulomb blockade emitter devices is limited by $\beta =
1/(1/\beta_1+1/\beta_2)\approx10^3$.

Purportedly, there is an indirect influence of the emitter operation on the
electron current. The Coulomb blockade of holes leads to an reduced hole
diffusion length as well. Thus, the emitter design can use thinner poly--Si
layers which reduces the emitter delay of the electrons in turn.

The down--scaled bipolar transistor might serve as a current multiplier or
readout device for Coulomb blockade circuits. It is well known that besides
many advantageous features single electron transistors display a low current
gain $\beta\approx1$, which, among others, limits their use for logic
applications. It can be envisioned that the transistor studied in this paper
can be used as a current multiplier. Given a single electron transistor output
current of $\sim1\,$nA, a current amplification of $\beta=10^3$ would lead to
much more manageable currents. For instance, MOSFET sensing operating with
voltage swings of $\Delta V\approx100\,$mV, would require a load resistance of
1\,k$\Omega$, which is thoroughly feasible.

{\em Conclusions --}\/ The performance of a small emitter window size bipolar
transistor is studied assuming Coulomb blockade effects in the poly--Si
emitter. A reduced hole injection current is found leading to an improved
current drive $\beta$. The transistor is discussed as readout device for
Coulomb blockade systems.

\bibliographystyle{apsrev}
\bibliography{bjt,set}

\end{document}